\documentclass[11pt]{article}

\begin{document}

\markboth{Edmundo M. Monte}
{Topological Properties from  Einstein's Equations}

%
%
\title{TOPOLOGICAL PROPERTIES FROM EINSTEIN'S EQUATIONS?}

\author{EDMUNDO M. MONTE\thanks{e-mail: edmundo@fisica.ufpb.br}\\
Departamento de F\'{\i}sica, Universidade \\Federal da
Para\'{\i}ba, 58059-970, Jo\~{a}o Pessoa, Para\'{\i}ba, Brasil}

\maketitle


\begin{abstract}
In this work we propose a new procedure for to extract global information of a space-time.
 We considered a space-time immersed in a higher dimensional space and we formulate the
 equations of Einstein through of the Frobenius conditions to immersion. Through of an
 algorithm and the implementation into algebraic computing system we calculate normal
  vectors from the immersion to find out the second fundamental form. We make a
  application  for space-time with spherical symmetry and static. We solve the equations
  of Einstein to the vacuum and we obtain space-times with different topologies.

\end{abstract}

\section{Introduction}

We are living a fertile period of attempts to solve great problems not solved of the physics.
 We can mention some of them, for instance in the cosmology area: What is the true cause of
 the accelerating expansion we seem to it observes now, and is it likely to it continues
 indefinitely into the future? What lack in the theory of the general relativity to
 explain this accelerated expansion in a consistent way? If the data astrophysicists
 are correct we needed to do a revision in the physics foundations, as for instance in
 the theory of the general relativity.

The Einstein's equations are local field equations describing local structure of
the space-time. The central idea in the formulation of the field equations is in the
question of how mass-energy generates curvature. One of requirement essential this
formulation is conservation of the mass-energy and it results the need of we define
 a geometric quantity that satisfies this condition. This quantity is the Einstein
 tensor that is composed as a combination of the Ricci's curvature and scalar of curvature,
 which are contractions of the tensor of Riemann. This last one is calculated with the metric.
 The solution of the equation of Einstein is a metric one that represents the gravitational
 field. Of this formulation we don't have information which is the shape of the space-time,
 solution of the field equations. From Einstein equations we have no global or topological
 information. The Einstein's  equations  being  partial  differential  equations,
they describe only local properties of space-time.  The reason is in the fact that when we
calculated the tensor of Riemann with the metric we cannot distinguish the shape of the
space-time. A trivial example is in the geometry of surfaces, where we don't get to
differentiate the shape between a plan and a cylinder from the Gaussian curvature.
 However, when we observed these surfaces immersed in $\Re^{3}$, we can distinguish
 the shape of the surfaces through the extrinsic curvature, when we used the equation
 of Gauss. This is a subtle idea that results in an 'expansion of the equations of Einstein',
 as we will see along this note.

The idea of extra dimensions at the space-time is very old, we can mention the work
of Kaluza, a theory that requests high energies. Today we are reviving this idea in
other approaches, as in the theory of strings, brane-world or even in other independent
 models of these where we considered a space-time immersed in a space of larger dimension,
 confinement of the gauge fields and the fact of only gravitons to escape for the extra
 dimensions, besides requesting energy in a 'low scale'.\cite{Kaluza}\cite{ME}

In this communication we propose a procedure where the space-times are immersed in higher
dimensional spaces, natural hypothesis requested for us to acquire topological information
of the space-time comings of the solution of the equations of Einstein. As example, we will
take two isometric immersions of a space-time with spherical symmetry and static in a
space pseudo-euclidean, in each case, we calculate the second fundamental form and through
the equation of Gauss we calculate the tensor of Einstein. Soon after we solve the equations
of Einstein to the vacuum we obtain space-times with different topologies.

\section{Deriving Schwarzschild Solution from Extrinsic Geometry}

Now we make the usual construction of the a metric with spherical symmetry and we solve
 the equations of field of the gravitation to vacuum  with static source and spherical symmetry.
 The solution is assumed to be spherically symmetric, static and into vacuum.
Simplifying the general metric with spherical symmetry, for the usual method to find out the metric,
\begin{equation}
ds^{2} = A(r)dt^{2} - B(r)dr^{2} - r^{2}(d\theta ^{2} + sin^{2}\theta d\phi ^{2})
\end{equation}

Consider two isometric immersions of a space-time with metric above from the isometric condition,\cite{Kobayashi}
\begin{equation}
g_{ij} = Y^\mu_{, i} Y^\nu_{, j} \eta_{\mu\nu},
\end{equation}
into a pseudo Euclidean manifold of six dimensions, with different signatures:\\

- The first immersion: \\
\[
ds^{2}=\;
dY_{1}^{2}+dY_{2}^{2}-dY_{3}^{2}-dY_{4}^{2}-dY_{5}^{2}-dY_{6}^{2}.
\]

-The second immersion: \\
\[
ds^{2}=\;{d{Y'}_{1}}^{2}-{d{Y'}_{2}}^{2}-{d{Y'}_{3}}^{2}-
{d{Y'}_{4}}^{2}-{d{Y'}_{5}}^{2}-{d{Y'}_{6}}^{2}.
\]
Respectively given by
\begin{equation}
\left\{
\begin{array}{l}
Y_{1}=r\mbox{sin}\theta \mbox{sin}\phi\\
\vspace{1mm} Y_{2}=r\mbox{sin}\theta \mbox{cos}\phi\\
\vspace{1mm}Y_{3}=r\mbox{cos}\theta\\
\vspace{1mm}
Y_{4}=\alpha(r)\\
\vspace{1mm}
Y_{5}=A(r)^{1/2}\mbox{cos}t\\
\vspace{1mm}Y_{6}=A(r)^{1/2}\mbox{sin}t\\
where
\;\;
d\alpha/dr=F(A(r),B(r))
\vspace{1mm}\end{array}\right.\;\;\;\;
\mbox{and}\;\;\;
\left\{
\begin{array}{l}
Y'_{1}=r\mbox{sin}\theta \mbox{sin}\phi\\
\vspace{1mm} Y'_{2}=r\mbox{sin}\theta \mbox{cos}\phi\\
\vspace{1mm}Y'_{3}=r\mbox{cos}\theta\\
\vspace{1mm}
Y'_{4}=\beta(r)\\
\vspace{1mm}
Y'_{5}=A(r)^{1/2}\mbox{cosh}t\\
\vspace{1mm}Y'_{6}=A(r)^{1/2}\mbox{sinh}t\\
where
\;\;
d\beta/dr=F(A(r),B(r))\\
\vspace{1mm}\end{array}\right.\label{eq:YY}
\end{equation}

Through  of  an  algorithm  and a implementation into algebraic computing system we calculate  normal vectors from the immersion to find
out  the  second fundamental  form,\cite{Kobayashi}\cite{Manfredo}
\[
b_{ijA} = Y^\mu_{; ij} N^\nu_A \eta_{\mu\nu}.
\]
We substitute those components in the equation of Gauss,\cite{Kobayashi}\cite{Manfredo}
\[
R_{nijk} = g^{AB}(b_{ikA}b_{jnB} - b_{ijA}b_{knB}),
\]
for to find the tensor of Riemann and soon after the tensor of Einstein $G_{ij}$.
We use the field equations $G_{ij}=0$ to find $A(r)$ and $B(r)$. We have been finding two relative metrics to the immersions mentioned above, $g$ and $g'$ metrics respectively:
\begin{equation}
ds^{2} = (1 - 2m/r)dt^{2} - (1 - 2m/r)^{-1}dr^{2} - r^{2}(d\theta ^{2} + sin^{2}\theta d\phi ^{2})
\end{equation}

\begin{equation}
ds^{2} = (32m^{3}/r) exp(-r/2m)(dv^{2} - du^{2}) - r^{2}(d\theta ^{2} + sin^{2}\theta d\phi ^{2}),
\end{equation}

where $u=u(Y'_6)$ and $v=v(Y'_5)$.

\section{Topological Properties from  Einstein's Equations}

The Schwarzschild's  solution representing the  empty
space-time with spherical symmetry  outside of a body with spherical
mass. Using spherical coordinates $(t,r,\theta ,\phi )$ this solution
is given by \cite{HE}
\begin{equation}
ds^{2} = (1 - 2m/r)dt^{2} - (1 - 2m/r)^{-1}dr^{2} - r^{2}(d\theta ^{2} + sin^{2}\theta d\phi ^{2})
\end{equation}
where  $ M =c^{2}mG^{-1}$, $c$ is the speed of light and  $G$ is the
gravitational constant, and it coincides with $(4)$. Notice that in these coordinates regions
 $r=0$ and $r=2m$ are  singular  and need to be removed.
When we remove the   surface $r=2m$, the manifold becomes separated in two
disconnected components, one for $2m<r<\infty $ and the other for $0<r<2m $.
Since we  are  dealing with the existence  of the  metric
associated to a  physical
space, we require a connected space. Therefore, we define the following regions:\cite{Oneill}\\
a)The exterior Schwarzschild space-time  $(V_{4},g)$:\\
$ V_{4}=P_{I}^{2}\times S^{2}\;$; $\; P_{I}^{2}=\{(t,r)\in \Re^{2}|\; r>2m\}\;$\\
b) The Schwarzschild black hole  $(B_{4},g)$:\\
$ B_{4}=P_{II}^{2}\times S^{2}\;$,
$\; P_{II}^{2}=\{(t,r)\in \Re^{2} |\; 0<r<2m\}\;$\\
In both cases, $S^{2}$  is the sphere of radius $r$ and  the metric $g$ is
given from $(6)$.

Consider two isometric immersions of space-time $(E,g)=([P_{I}^{2}\cup P_{II}^{2}]\times S^{2},g)$,
into a pseudo Euclidean manifold of six dimensions, with different signatures:\\

- The Kasner immersion: \cite{Kasner}\\
\[
ds^{2}=\;
dY_{1}^{2}+dY_{2}^{2}-dY_{3}^{2}-dY_{4}^{2}-dY_{5}^{2}-dY_{6}^{2}.
\]

-The Fronsdal immersion: \cite{Fronsdal}\\
\[
ds^{2}=\;{d{Y'}_{1}}^{2}-{d{Y'}_{2}}^{2}-{d{Y'}_{3}}^{2}-
{d{Y'}_{4}}^{2}-{d{Y'}_{5}}^{2}-{d{Y'}_{6}}^{2},
\]
Respectively given by
\begin{equation}
\left\{
\begin{array}{l}
Y_{1}=r\mbox{sin}\theta \mbox{sin}\phi\\
\vspace{1mm} Y_{2}=r\mbox{sin}\theta \mbox{cos}\phi\\
\vspace{1mm}Y_{3}=r\mbox{cos}\theta\\
\vspace{1mm}
Y_{4}=\alpha(r)\\
\vspace{1mm}
Y_{5}=(1-2m/r)^{1/2}\mbox{cos}t\\
\vspace{1mm}Y_{6}=(1-2m/r)^{1/2}\mbox{sin}t\\
 where
\;\;
(d\alpha/dr)^{2}=\frac{2mr^{3}+m^2}{r^{4}-2mr^{3}}\\
\vspace{1mm}\end{array}\right.\;\;\;\;
\mbox{and}\;\;\;\left\{
\begin{array}{l}
Y'_{1}=r\mbox{sin}\theta \mbox{sin}\phi\\
\vspace{1mm} Y'_{2}=r\mbox{sin}\theta \mbox{cos}\phi\\
\vspace{1mm}Y'_{3}=r\mbox{cos}\theta\\
\vspace{1mm}
Y'_{4}=\beta(r)\\
\vspace{1mm}
Y'_{5}=(1-2m/r)^{1/2}\mbox{cosh}t\\
\vspace{1mm}Y'_{6}=(1-2m/r)^{1/2}\mbox{sinh}t\\
 where
\;\;
(d\beta/dr)^{2}=\frac{2mr^{3}-m^2}{r^{4}-2mr^{3}}\\
\vspace{1mm}\end{array}\right.\label{eq:YY}
\end{equation}

In order to determine the metric $g'$
(extension of $g$), define the  new coordinates $u$  and $v$  by:\cite{EM}\\

- For $r>2m$,
\begin{equation}
v=\frac{1}{4m}(\frac{r}{2m})^{1/2}exp(\frac{r}{4m}){Y'}_{5}\; \; \mbox{and}
\;\;u=\frac{1}{4m}(\frac{r}{2m})^{1/2}exp(\frac{r}{4m}){Y'}_{6}.
\end{equation}
- For $0<r<2m$,
\begin{equation} v=\frac{1}{4m}(\frac{-r}{2m})^{1/2}exp(\frac{r}{4m}){Y'}_{5}\;
\; \mbox{and}
\;\;u=\frac{1}{4m}(\frac{-r}{2m})^{1/2}exp(\frac{r}{4m}){Y'}_{6},
\end{equation}
where
\begin{equation}
u^{2} - v^{2}=(\frac{r}{2m} - 1)exp(\frac{r}{2m})\;\Longleftrightarrow
\;{Y'}_{6} ^{2} - {Y'}_{5}^{2}=16m^{2}(1 -\frac{2m}{r}).
\end{equation}
Now $r=r({Y'}_{5},{Y'}_{6})$ is
implicitly defined by equation $(10)$, while $t=t({Y'}_{5},{Y'}_{6})$ is
implicitly defined by
\begin{equation}
{Y'}_{5}/{Y'}_{6}=tgh(\frac{t}{4m}).
\end{equation}

Now we can write the $g'$ metric,

\begin{equation}
ds^{2} = (32m^{3}/r) exp(-r/2m)(dv^{2} -
du^{2}) - r^{2}(d\theta ^{2} +
sin^{2}\theta d\phi ^{2}).
\end{equation}

We notice that $(E',g')$ is the extension of
$(E,g)$, it calculated through of the coordinates of immersion and it coincides with $(5)$.
The space $(E',g')$ don't has a singularity at
$r=2m$. We know that $(E,g)$ is disconnected because it is composed by two
connected components. When we calculated the extension $(E',g')$ through the
Fronsdal  immersion we see that it is connected.

We use the isometric immersion formalism to establish the extension of
$(E,g)=([P_{I}^{2}\cup P_{II}^{2}]\times S^{2},g)$, denoted  by
$(E',g')=(Q^{2}\times S^{2},g')$, where  $Q^{2}$  is the Kruskal plane.\cite{Kruskal}
The topology of a gravitational field outside of a body with spherical symmetry is given by
$\Re^{2}\times S^{2}$. We have that the topology of
$(E,g)$ is given by $(\Re^{2}-\{(t,r)\in \Re^{2} |\;r=2m\})\times S^{2}$, is different
from the topology $\Re^{2}\times S^{2}$  of $(E',g')$.\cite{EM}

\section{Conclusion}

We start of one space-time immersed in a space pseudo-euclidean. We compose the tensor of Riemann through of the Frobenius conditions to immersion. In particular we solved the field equations for the vacuum and we assume a space-time with spherical symmetry and static. We found the immersion coordinates and later we calculate the second fundamental form through an algorithm using algebraic computation. Soon after we calculate the tensor of Riemann and Einstein through the equation of Gauss. We found the solutions of the equations of Einstein for the vacuum: the metric of Schwarzschild in the usual coordinates and Kruskal metric in this approach. Finally we found the topological characteristics associated the those solutions. We conclude that is possible to extract global information of the solutions of the equations of Einstein with this new procedure. This procedure results a source of global information to the gravitation field, not supplied by the solutions of the Einstein equations of the general relativity. We suggest that those additional information to gravitational field  brought from of the geometric entities of the immersion can be associated to physical amounts that help, for instance, to solve current problems in cosmology.

\section{Acknowledgments}
I would like to thank to Professor Marcos Maia for introducing me in this research area and for many discussions happened along the years regarding this theme. Thanks also to CAPES-Brazil (grant- 3749/09-6-senior stage-Spain) for partial financial support.

\end{document}